\documentclass[%
 reprint,
 amsmath,amssymb,
 aps,
]{revtex4-2}
\usepackage{float}
\usepackage{graphicx}
\usepackage{dcolumn}
\usepackage{bm}
\usepackage{color}
\newcommand{\vc}{\mathbf}
\usepackage{array}
\newcolumntype{P}[1]{>{\centering\arraybackslash}p{#1}}

\begin{document}

\preprint{APS/123-QED}

\title{Bulk Viscosity of the Rigid Rotor One-Component Plasma}

\author{Jarett LeVan}
\author{Marco D.~Acciarri}
\author{Scott D.~Baalrud}
\affiliation{%
Department of Nuclear Engineering and Radiological Sciences, University of Michigan, Ann Arbor, MI 48109, USA
}%

\date{\today}

\begin{abstract}
Bulk viscosity of a plasma consisting of strongly coupled diatomic ions is computed using molecular dynamics simulations. The simulations are based on the rigid rotor one-component plasma, which is introduced as a model system that adds two degrees of molecular rotation to the traditional one-component plasma. It is characterized by two parameters: the Coulomb coupling parameter, $\Gamma$, and the bond length parameter, $\Omega$. Results show that the long-range nature of the Coulomb potential can lead to long rotational relaxation times, which in turn yield large values for bulk viscosity. The bulk-to-shear viscosity ratio is found to span from small to large values depending on the values of $\Gamma$ and $\Omega$. 
Although bulk viscosity is often neglected in plasma modeling, these results motivate that it can be large in molecular plasmas with rotational degrees of freedom. 
\end{abstract}

\maketitle

\section{Introduction}
Bulk viscosity is a transport coefficient associated with the irreversible resistance to expansion or compression of a fluid~\cite{landau2013fluid}. The resistance arises from time-lags of energy transfer between translational modes and the rest of the system (rotational modes, vibrational modes, internal energy, etc.). For more than a century following its appearance in Stokes' 1845 paper on viscous fluids, the very existence of bulk viscosity was the source of considerable controversy~\cite{Stokes_1845, Truesdell_1954, Malbrunot_1983, Emanuel_1990, GadelHak_1995}. However, it has become clear in recent decades that fluids can exhibit large values of bulk viscosity, especially when they are composed of molecules~\cite{Tisza_1942, Prangsma_1973, Guo_2001, Pan_2004, Cramer_2012, Sharma_2022, Sharma_2023}. A number of studies have investigated the effects of bulk viscosity in neutral gases and showed it can significantly alter acoustic attenuation~\cite{Hansen_2013}, shock wave structure~\cite{Emanuel_1994, Elizarova_2007, Kosuge_2018}, turbulence~\cite{Pan_Johnsen_2017,Touber_2019, Chen_2019}, and instabilities~\cite{Sengupta_2016, Singh_2021}. 

Although similar fluid phenomena are also present in plasmas, little attention has been paid to bulk viscosity in plasma physics. 
Perhaps this is related to an early study, which used molecular dynamics simulations to show that the bulk viscosity of the one-component plasma is small in comparison to the shear viscosity \cite{Hansen_1975}.
This is the case in dilute monatomic neutral fluids as well \cite{Nettleton_1958,Sharma_monatomic}. 
However, as in neutral fluids, bulk viscosity may be much larger if internal degrees of freedom such as vibration or rotation are present~\cite{Herzfeld_1928, Tisza_1942, Cramer_2012}. 
This motivates revisiting the calculation of bulk viscosity in the context of molecular plasmas. 
Such a study is timely, as recent papers have begun to explore the potential implications of bulk viscosity in plasmas. Istomin \textit{et al}~\cite{Istomin_2017} demonstrated that electronic excitation can significantly increase the bulk viscosity in ionized gases, potentially impacting shocks in spacecraft re-entry. In a plasma astrophysics context, Beattie \textit{et al}~\cite{beattie_2023} showed that bulk viscosity can strongly suppress compressible modes in the turbulent dynamo. And in studying the propagation of magnetoacoustic waves using magnetohydrodynamics, Cunha \textit{et al}~\cite{Cunha_2024} showed that bulk viscosity can be a dominant source of energy dissipation in expanding plasma flows, and that the rate of dissipation can be controlled by tuning the intensity and orientation of an applied magnetic field. These papers demonstrate that bulk viscosity can substantially alter the macroscopic fluid dynamics of plasmas. Therefore, it is important to characterize the bulk viscosity in a molecular plasma for the sake of more accurate fluid models. 

In this work, we compute the coefficient of bulk viscosity for a system of strongly coupled diatomic ions using molecular dynamics (MD) simulations. We first establish the rigid rotor one-component plasma (ROCP) model, which consists of diatomic molecules with a neutral atom bonded to an ion at a fixed bond length $r_\textrm{B}$. It is a variation of the traditional one-component plasma (OCP), which is commonly used to study strongly coupled monatomic ions \cite{Baus-Hansen-1980}. The properties of the traditional OCP are characterized by just one dimensionless parameter, the Coulomb coupling parameter 
\begin{equation}
\label{eq:gamma}
    \Gamma = \frac{q^2/a}{4 \pi \epsilon_0 k_\textrm{B} T} ,
\end{equation} 
where $a = (3/4\pi n)^{1/3}$ is the average inter-particle spacing and $T$ is the temperature. 
In contrast, characterizing the ROCP requires two parameters due to the rotational degrees of freedom. 
In addition to $\Gamma$, which is associated with translational degrees of freedom, the bond length parameter 
\begin{equation}
\label{eq:omega}
    \Omega = \frac{r_\textrm{B}}{a} ,
\end{equation}
is associated with the rotational degrees of freedom. 
Together, $\Omega$ and $\Gamma$ fully characterize the ROCP. 

Under this framework, equilibrium MD simulations were run to compute both bulk ($\eta_v$) and shear ($\eta$) viscosity using the Green-Kubo formalism \cite{Green_1952,Kubo_1957,Hansen_2013}. Results are compared with those for the OCP model. 
It is found that the shear viscosity of the OCP and ROCP are identical, but the bulk viscosity is much larger in the ROCP than in the OCP. 
For the OCP, bulk viscosity is at least an order of magnitude smaller than shear viscosity at any $\Gamma$ value. 
For the ROCP, bulk viscosity can exceed the shear viscosity by several orders of magnitude at particular combinations of $\Gamma$ and $\Omega$ values. 
Large values of bulk viscosity are found to be associated with a long time decay of the pressure autocorrelation function, which can be traced to a long relaxation time between the translational and rotational degrees of freedom. 
In particular, it is found that small values for $\Gamma$ and $\Omega$ lead to larger $\eta_v$.
It is thought that this is because the long-range nature of the Coulomb force effectively shields the rotational degree of freedom, and that the shielding is more effective at smaller values of the bond length compared to interparticle spacing ($\Omega$) and smaller values of the Coulomb coupling strength ($\Gamma$). 
In this regime, any deviation from equilibrium will cause energy to be temporarily trapped in rotational degrees of freedom, leaving translational energy away from it's equilibrium value for a long time. 
As an example, at a density corresponding to STP ($n = 2.5 \times 10^{25}~\textrm{m}^{-3}$),  room temperature ($T=293$~K), and a bond length corresponding to N$_2$ ($r_\textrm{B} = 1.09~\mathrm{\AA}$), or in dimensionless units $\Gamma = 31.5$ and $\Omega = 0.05$, we predict a bulk to shear ratio for the ROCP of $\eta_v / \eta \sim 10^3$. To contrast, at these same conditions we found the OCP has $\eta_v / \eta \sim 10^{-2}$, and Sharma \textit{et al}~\cite{Sharma_2022} found neutral N$_2$ to have $\eta_v / \eta \sim 1$.
This demonstrates that the presence of molecular ions can dramatically impact bulk viscosity.  

Previous models from neutral gas dynamics have been developed to connect the rotational relaxation time to the bulk viscosity. 
Here, we tested the application of Kustova's formula to plasmas by computing the rotational relaxation time from separate MD simulations, using the result in Kustova's formula, and comparing the predicted bulk viscosity coefficient with the result of the Green-Kubo calculation. 
This comparison showed excellent agreement over the range of parameters that both methods could be evaluated. 
Due to the long time required for convergence of the Green-Kubo relations, values could only be computed for a limited range of $\Gamma$ and $\Omega$ from this method; $\Gamma$ in the strong coupling regime from 1-100, and $\Omega$ from 0.1 - 0.3. To extend this, rotational relaxation data was acquired for a wider range of $\Omega$ values (0.01 - 0.30), and the results input into Kustova's formula in order to predict values for bulk viscosity across a larger parameter space. 
Furthermore, a model for the relaxation time is applied to provide a practical formula for estimating the ROCP bulk viscosity. 

This paper is organized as follows: Section~\ref{sec:rocp} introduces the ROCP model. Section~\ref{sec:bv} introduces some historical context on bulk viscosity, describes previous models connecting bulk viscosity in molecular gases with the rotational relaxation time, where bulk viscosity arises in hydrodynamics, and how it can be computed from the Green-Kubo relations. 
Section~\ref{sec-bulk} describes the MD simulation setup and results of the viscosity coefficients for the OCP and ROCP computed from the Green-Kubo relations. 
Section~\ref{sec-rot} describes an MD setup to compute the rotational relaxation time, and a test of the model connecting relaxation time to bulk viscosity, along with a practical formula for computing the bulk viscosity of the ROCP. 
Finally, some potential implications of bulk viscosity to sound attenuation, shocks, and turbulence are discussed along with concluding comments in Sec.~\ref{sec:conclusion}. 

\section{The Rigid Rotor One-Component Plasma\label{sec:rocp}}
The one-component plasma (OCP) model is a well-established means for studying the properties of strongly coupled particles~\cite{Baus-Hansen-1980}. When applied to plasmas, ions are often modelled as point particles, each with charge $q$ and mass $m$, and collectively, a density $n$ and temperature $T$. Electrons are not modelled directly, but taken to provide a non-interacting, charge neutralizing background. The OCP applies particularly well to systems where the electrons are at a weaker coupling strength than the ions, such as when the electron temperature is much larger than the ion temperature. A particularly attractive feature of the OCP is that when time is quantified in dimensionless units of the plasma period $\omega_{p}^{-1} = (\epsilon_o m/q^2n)^{1/2}$ and space in units of the interparticle spacing $a=(3/4\pi n)^{1/3}$,  
it is entirely characterized by the Coulomb coupling parameter $\Gamma$, defined in Eq.~(\ref{eq:gamma}). 

By treating ions as point particles, the traditional OCP is limited to describing monatomic ions. However, there are many examples of plasmas in which strongly coupled molecular ions exist. Atmospheric pressure plasmas commonly reach ion densities sufficient for strong coupling \cite{Acciarri_2022, Lo_2017, Minesi_2020, van_der_Horst_2012} and can be composed, in large part, of molecular ions \cite{Fu_2018, Farouk_2006, Balcon_2008}. Additionally, in ultracold plasmas strong coupling has been achieved and studied with NO$^+$ ions \cite{Morrison_2008, Sadeghi_2011,Sadeghi_2014}. 
And in high energy density plasmas, strongly coupled molecular ion species may form during implosions, since molecular species are common in dense shell material and implosions often reach strong coupling \cite{Hu_2010, Hurricane_2023}.

To study the dynamics of strongly coupled diatomic ions, we establish the rigid rotor one-component plasma (ROCP). In this model, diatomic ions consist of two point particles, each with mass $m$, rigidly bonded to each other at some fixed distance, $r_\textrm{B}$. One particle represents an atomic ion with charge $q$, that interacts with other ions through the Coulomb potential $V_c(r) = q^2/(4\pi \epsilon_0 r)$. The other represents a neutral atom which does not interact with the rest of the system. 
The justification for the neutral atom to be passive (interacting only through its bonding to an ion) is that the Coulomb potential of the ion is much longer range than the interatomic potential of the neutral atom, and therefore shields it in any interaction. 
This expectation that ion-neutral and neutral-neutral interactions are very weak compared to the Coulomb interaction is tested and validated in Sec.~\ref{sec-rot}. 

This model behaves in much the same way as the traditional OCP, but with the addition of two rotational degrees of freedom. Therefore, some properties, like shear viscosity, remain unchanged; see Sec.~\ref{sec-bulk}. However, the rotational degrees of freedom add an associated relaxation time, $\tau_{\textrm{rot}}$, which impacts the bulk viscosity $\eta_v$. Additionally, the coupling parameter $\Gamma$ cannot fully characterize this model like it does the traditional OCP. 
The additional rotational degrees of freedom are quantified by the dimensionless bond length parameter $\Omega$, defined in Eq.~(\ref{eq:omega}). 

Physically, $\Omega$ is the ratio of a molecule's bond length to the average inter-particle spacing. It quantifies how important rotation is in collisional processes. Larger values of $\Omega$ correspond to larger changes in the force on an ion as it travels around its circumference of rotation. Conversely, at sufficiently small $\Omega$, molecules begin to act like atoms, so rotation has little effect on collisions. 

The rotational coupling parameter can fully account for the effects of molecular rotation in this framework. 
Consider the torque equation for a given rigid rotor
\begin{equation}
    I\frac{d\bm{\omega}}{dt} = \bm{r} \times \bm{F}
    \label{eq-torque}
\end{equation}
where $I = \frac{1}{2}mr_\textrm{B}^2$ is the moment of inertia, $\bm{\omega}$ is the angular velocity, $\bm{r}$ points from the center of mass to the ion, and $\bm{F} = q^2/(4\pi \epsilon_0) \sum_{j}\bm{r}_{j}/r_{j}^3$ is the Coulomb force on the ion due to the other ions in the system. 
By using dimensionless units of time in terms of $\omega_p^{-1}$ and length in terms of $a$
\begin{equation}
\tilde{\bm{x}} = \bm{x}/a, \ \ \ \tilde{t} = t\omega_p,  
\end{equation}
Eq.~(\ref{eq-torque}) can be expressed in terms of $\Omega$
\begin{equation}
    \frac{d\tilde{\omega}}{d\tilde{t}} = \frac{1}{3} \Omega^{-1} \sum_{j}\frac{\sin\theta_j}{\tilde{r}_{j}^2},
\end{equation}
where $\theta_j$ is the angle between $\bm{r}$ and $\bm{r}_j$.
Similarly, since only one particle in the rigid body interacts, the translational equation of motion looks the same as the traditional OCP 
\begin{equation}
    \frac{d\tilde{\bm{v}}}{d\tilde{t}} = \frac{1}{3} \sum_{j}\frac{\tilde{\bm{r}}_{j}}{\tilde{r}_{j}^3}.
\end{equation}
The dependence on $\Gamma$ arises from the initial condition, specifically by setting the particle velocities corresponding to a set temperature $\tilde{v}_o = \tilde{v}_T = \sqrt{2/(3 \Gamma)}$. As with $\Gamma$ for the OCP, since the dimensionless equations of motion of the ROCP can be written in terms of $\Gamma$ and $\Omega$, all properties can be characterized by these parameters when cast in the associated dimensionless units. 
One note when comparing the OCP and ROCP is that the mass ($m$) is the atomic mass in the OCP, but the molecular mass in the ROCP. 

\section{Overview of Bulk Viscosity \label{sec:bv}}
\subsection{Underlying Mechanisms}
In his original 1845 paper on viscous fluids, Stokes assumed that a fluid, upon being subjected to a given dilatation, will still have a mechanical scalar pressure given by its thermodynamic equation of state, i.e., $p_m = p(n, T)$~\cite{Stokes_1845}. This assumption, known as Stokes' hypothesis, implies that the coefficient of bulk viscosity ($\eta_v$) is zero. By definition, $p_m = -\frac{1}{3}\textrm{Tr}\lbrace \bm{\Pi} \rbrace$, where $\bm{\Pi}$ is the stress tensor, which for a compressible Newtonian fluid takes the form
\begin{equation}
    \bm{\Pi} = -p_t \vc{I} + \eta \biggl[ \nabla \vc{V} + (\nabla \vc{V})^T - \frac{2}{3} (\nabla \cdot \vc{V}) \vc{I} \biggr] + \eta_v (\nabla \cdot \vc{V}) \vc{I}
\end{equation}
where $p_t = p(n, T)$ is the thermodynamic pressure and $\vc{V}$ is the fluid velocity, yielding the result
\begin{equation}
    p_{m} - p_{t} = -\eta_v \nabla \cdot \vc{V}.
    \label{eq-bulk-pressure}
\end{equation}
Clearly, if $p_{m} = p_{t}$, then $\eta_v = 0$.

Some 80 years after Stokes' publication, Herzfeld and Rice proposed a mechanism by which this assumption could be violated \cite{Herzfeld_1928}. If a fluid is composed of molecules, not atoms, then a given expansion or compression will cause a temporary violation of the equipartition theorem, as work is done only at the expense of translational kinetic energy. It takes a finite amount of time for an equilibrium between degrees of freedom to be restored, during which $p_t \neq p_m$, implying a finite value for $\eta_v$.

Based on this mechanism, Tisza, in 1941, derived the following formula for bulk viscosity
\begin{equation}
    \eta_v = \rho a_s^2 \frac{\gamma - 1}{\gamma} \sum_i \frac{c_{v,i}}{c_v} \tau_i
    \label{eq-tisza}
\end{equation}
where $\rho$ is the mass density, $a_s$ is the speed of sound in the absence of viscosity, $\gamma$ is the ratio of specific heats, $c_v$ is the specific heat at equilibrium, $c_{v,i}$ is the specific heat of the $i^{\textrm{th}}$ vibrational mode, and $\tau_i$ is the relaxation time of the $i^{\textrm{th}}$ vibrational mode \cite{Tisza_1942}. Importantly, Tisza assumed molecular rotation to be unimportant because rotational relaxation is often a much faster than vibrational relaxation. This formula has since served as a standard means for computing the bulk viscosity of neutral fluids \cite{Cramer_2012}. Tisza applied the appropriate values for CO$_2$ at STP conditions and predicted a bulk to shear viscosity ratio of $\eta_v / \eta > 1000$. 

In 2019, Kustova \textit{et al}~\cite{Kustova_2019} derived a more general expression for bulk viscosity from kinetic theory without neglecting molecular rotation. They found
\begin{equation}
    \eta_v = nk_\textrm{B} T R \left( \frac{c_\textrm{int}}{c_v}\right)^2 \left( \frac{c_{\textrm{rot}}}{\tau_{\textrm{rot}}} + \frac{c_{\textrm{vib}}}{\tau_{\textrm{vib}}}\right)^{-1}
    \label{eq-kustova}
\end{equation}
where $R$ is the specific gas constant, $c_\textrm{int}$ is the specific heat capacity of internal degrees of freedom, $c_\textrm{rot} = (f_\textrm{rot}/2)R$ is the specific heat capacity of rotational degrees of freedom, $c_\textrm{vib}$ is the specific heat capacity of vibrational degrees of freedom, $\tau_{\textrm{rot}}$ is the rotational relaxation time, and $\tau_{\textrm{vib}}$ is the vibrational relaxation time. 
Here, $f_\textrm{rot}$ is the number of rotational degrees of freedom. 
In Kustova's formula, the coefficient of bulk viscosity is set by the faster of the two relaxation processes. As a result, they predict CO$_2$ to have a bulk to shear ratio on the order of 1 at STP conditions~\cite{Kustova_2019}. 

In the ROCP, molecules do not vibrate. 
In the context of the physical systems of interest, rotational relaxation is, in almost all cases, significantly faster than vibrational relaxation. As such, Eq.~(\ref{eq-kustova}) predicts that rotation will set the coefficient of bulk viscosity, rather than vibration, justifying the application of the ROCP model. 
For the ROCP, $c_\textrm{int} = c_\textrm{rot} = R$, $c_\textrm{vib} = 0$, and $c_v = R(\frac{5}{2}+ c_{v,\textrm{ex}})$. 
Here, the factor of $5/2$ comes from the 5 total degrees of freedom of molecular motion (3 translational and 2 rotational), and $c_{v,\textrm{ex}}$ is the excess (non-ideal) specific heat at constant volume. With these substitutions, Eq.~(\ref{eq-kustova}) becomes
\begin{equation}
    \eta_v = \left( \frac{5}{2} + c_{v,\textrm{ex}}\right)^{-2} n k_\textrm{B} T   \tau_{\textrm{rot}}.
    \label{eq-simple}
\end{equation}
Since $c_{v,\textrm{ex}}$ is associated with the molecular configuration, it is not expected to be significantly influenced by the rotational  degrees of freedom. 
Therefore, it can be modeled using a fit to Hansen's data for excess specific heat of the OCP~\cite{Hansen_1973}: $c_{v,\textrm{ex}} = 0.00000286 \Gamma^3 - 0.00057249\Gamma^2 +  0.04278796\Gamma + 0.14948573$.
It will be shown below that Eq.~(\ref{eq-simple}) provides an accurate model for the bulk viscosity of the ROCP when an accurate model for $\tau_\textrm{rot}$ is provided. 

There exist several mechanisms outside the Herzfeld and Rice mechanism for bulk viscosity. In 1948, Hall proposed a mechanism known as intrinsic bulk viscosity, applicable to both atomic and molecular fluids~\cite{Hall_1948}. A collection of particles, upon being compressed, will not be compressed into the lowest internal energy configuration immediately. Thus, over the course of some finite amount of time, the particles will undergo a process of structural rearrangement, leading to a finite value for bulk viscosity. 
Intrinsic bulk viscosity is often small because structural rearrangement is a fast process that occurs on the timescale of translational motion. Nevertheless, it is important because this mechanism causes even monatomic gases to have finite bulk viscosity~\cite{Nettleton_1958, Sharma_monatomic}. In the traditional OCP, intrinsic bulk viscsosity is the only mechanism present. Previous work has shown the bulk viscosity of the OCP to be small compared to the shear viscosity \cite{Hansen_1975, Scheiner_2020}. 
Here, we obtain better resolved data that agrees with this conclusion; the bulk to shear visocsity ratio is found to be smaller than $0.1$ for all $\Gamma$ values. 
Therefore, the Herzfeld and Rice mechanism is expected to dominate in the ROCP.

Other origins of bulk viscosity that have been explored include a chemical non-equilibrium induced by nuclear reactions~\cite{Dong_2007} and electronic excitation~\cite{Istomin_2017}. However, these processes are not present in the ROCP and will not be discussed here.

\subsection{Hydrodynamic Equations \label{sec-hydro}}
Bulk viscosity affects the stress tensor $\vc{\Pi}$, which appears in the hydrodynamic equation for momentum conservation as 
\begin{equation}
    \rho \frac{d\vc{V}}{dt} = \nabla \cdot \bm{\Pi}
\end{equation}
where $d/dt = \partial/\partial t + \vc{V} \cdot \nabla$ is the convective derivative and $\rho$ is the mass density. It also influences the energy conservation equation as
\begin{equation}
    \frac{du}{dt} = \lambda \nabla^2 T + \bm{\Pi} : \nabla \vc{V}
\end{equation}
where $u$ is the energy density, $\lambda$ is the thermal conductivity, and $T$ is the temperature.
When combined with the mass conservation equation
\begin{equation}
    \frac{d\rho}{dt} = -\nabla \cdot (\rho \vc{V}),
\end{equation}
these form the complete set of Navier-Stokes hydrodynamic equations~\cite{Hansen_2013}. 

Bulk viscosity is a powerful transport coefficient in that it can, in principle, account for the dynamical effects of structural rearrangement, rotational relaxation, and vibrational relaxation processes with a single number. This can dramatically improve computational costs for fluid simulations, as these processes would otherwise need to be accounted for with additional rate equations \cite{Graves_1999}. However, it should be stressed that transport coefficients can only be utilized if the underlying relaxation process that they represent occur on a timescale much faster than the flow timescale, i.e., the condition of local thermodynamic equilibrium must be satisfied \cite{thompson1971compressible}. Molecular vibration often violates this condition, as vibrational temperatures can be highly elevated and the associated relaxation time very long. For this reason, it is common practice to only consider the contribution of molecular rotation to bulk viscosity \cite{Nagnibeda2009}. This serves as another justification for ignoring vibration in the ROCP.  

\subsection{Green-Kubo Relations}
The Green-Kubo relations relate linear transport coefficients to the integral of a macroscopic variable's equilibrium time autocorrelation function. For shear and bulk viscosity, they take the following forms \cite{Hansen_2013}
\begin{subequations}
\begin{align}
    \eta &= \frac{\mathcal{V}}{6 k_\textrm{B} T} \sum_{i=1}^3 \sum_{\stackrel{j=1}{j\neq i}}^3 \int_0^\infty dt \left< \Pi_{ij}(0) \Pi_{ij}(t)\right> ,    \label{eq-gk-shear}\\
    \eta_v &= \frac{\mathcal{V}}{k_\textrm{B} T} \int_0^\infty dt \left< \delta p(0) \delta p(t)\right>   
    \label{eq-gk-bulk}
\end{align}
\end{subequations}
where $\mathcal{V}$ is the volume, $\Pi_{ij}$ is the stress tensor, $\delta p = p_m - p_{t}$, and the angle brackets $\left< \ldots \right>$ represent an equilibrium ensemble average.
Here, $p_m(t) = -\frac{1}{3}\textrm{Tr}\lbrace \Pi_{ij} \rbrace$ is the mechanical pressure and the thermodynamic pressure $p_{t}$ is obtained from a time average of $p_m$ over the entire simulation. 
One should note that in accordance with Eq.~(\ref{eq-bulk-pressure}), $\delta p = -\eta_v \nabla \cdot \vc{V}$, leaving Eq.~(\ref{eq-gk-bulk}) as a rather intuitive statement. The Green-Kubo relations show that relaxation processes resulting from equilibrium fluctuations are equivalent to the linear response to an external perturbation.

For the ROCP, the stress tensor is related to the particle trajectories by
\begin{equation}
    \Pi_{ij} = -\frac{1}{\mathcal{V}} \sum_{\alpha=0}^{N_\alpha} \left[ m_\alpha v_{\alpha_i}v_{\alpha_j} + \sum_{\beta>\alpha}^{N_\alpha} (r_{\alpha \beta_i} F_{\alpha \beta_j}) + r_{\alpha,i} F_{C,j}\right]
    \label{eq:md-stress}
\end{equation}
where $N_\alpha$ is the number of atoms in the system, $m_\alpha$ is the mass of atom $\alpha$, $r_{\alpha_i}$ and $v_{\alpha_i}$ are the position and velocity of atom $\alpha$ in direction $i$, $F_{\alpha \beta}$ is the Coulomb force between atoms $\alpha$ and $\beta$, and $F_C$ is the constraint force on atom $\alpha$ resulting from a rigid bond. It is clearly seen that there are three contributions to the system's stress: a kinetic contribution resulting from each atom's kinetic energy, a potential term resulting from the pairwise interaction between atoms, and a bond contribution resulting from the implied force it takes to hold the distance between two bonded atoms constant. 


The Green-Kubo relations can be employed to compute transport coefficients from MD simulations~\cite{Hansen_1975,Saigo_2002,Donko_2008, Daligault_2014, Scheiner_2020}. To do so, the autocorrelation function must be discretized in time and cutoff at some long time $L$, by which the autocorrelation has decayed to zero and its integral has converged. In practice, MD simulations have a finite number of particles, so a single autocorrelation function is far too noisy for convergence to be reached. However, if a time-series $\Pi_{ij}(t)$ is generated from MD with length $t_N \gg L$, then $t_N - L$ separate autocorrelation functions can be computed, as each timestep (from $0$ to $t_N-L$) can be used as $t = 0$ in a new autocorrelation function. By averaging these separate autocorrelation functions together, one can obtain a smooth decay to zero, and hence a convergent viscosity integral. 
With these changes, Eqs.~(\ref{eq-gk-shear}) and (\ref{eq-gk-bulk}) become 
\begin{subequations}
\begin{align}
    \eta (t) &= \frac{\mathcal{V}}{6k_\textrm{B} T} \frac{\Delta t}{t_N - L + 1} \times \\ \nonumber
    &\sum_{\tau_t=0}^t \sum_{i=1}^3 \sum_{\stackrel{j=1}{j\neq i}}^3\sum_{\tau = 0}^{t_N - L}   \Pi_{ij}(\tau) \Pi_{ij}(\tau + \tau_t) ,
    \\
    \eta_v (t) &= \frac{\mathcal{V}}{k_\textrm{B} T} \frac{\Delta t}{t_N - L + 1} \sum_{\tau_t=0}^t \sum_{\tau = 0}^{t_N - L}  \delta p(\tau) \delta p(\tau + \tau_t) ,
    \label{eq-bulk-gk-sim}
\end{align}
\end{subequations}
where $\Delta t$ is the length of a timestep, $t_N$ is the total number of timesteps, and $L$ is the length of the autocorrelation function.

\section{MD Simulations and Results\label{sec-bulk}}
\subsection{MD Setup}
Equilibrium MD simulations were run to compute bulk and shear viscosity using the Green-Kubo relation from Eqs.~(\ref{eq-gk-shear}) and (\ref{eq-gk-bulk}). Simulations consisted of 20,000 diatomic ions in a box with periodic boundary conditions. Ions were modelled as rigid rotors with just one of the atoms ionized. Short-range interactions ($r < 5a$) were computed exactly using the Coulomb potential while long-range interactions ($r > 5a$) were computed using the particle-particle-particle-mesh (P$^3$M) algorithm as described in Ref.~\onlinecite{Hockney1966ComputerSU}. A timestep of $\Delta t = 0.01 \omega_p^{-1}$ was found to be sufficiently small for conserving energy, matching the timestep requirement for the OCP. To start the simulations, translational and rotational degrees of freedom were thermostat to the same temperature until equilibrium was reached. 
The system was subsequently evolved in the microcanonical (NVE) ensemble. The stress tensor was output every 10 timesteps and computed using Eq.~(\ref{eq:md-stress}), with a slight modification of the potential term due to the use of P$^3$M, which is discussed in detail in Ref.~\onlinecite{Daligault_2014}.

\begin{figure}[H]
    \centering
    \includegraphics[width=0.48\textwidth]{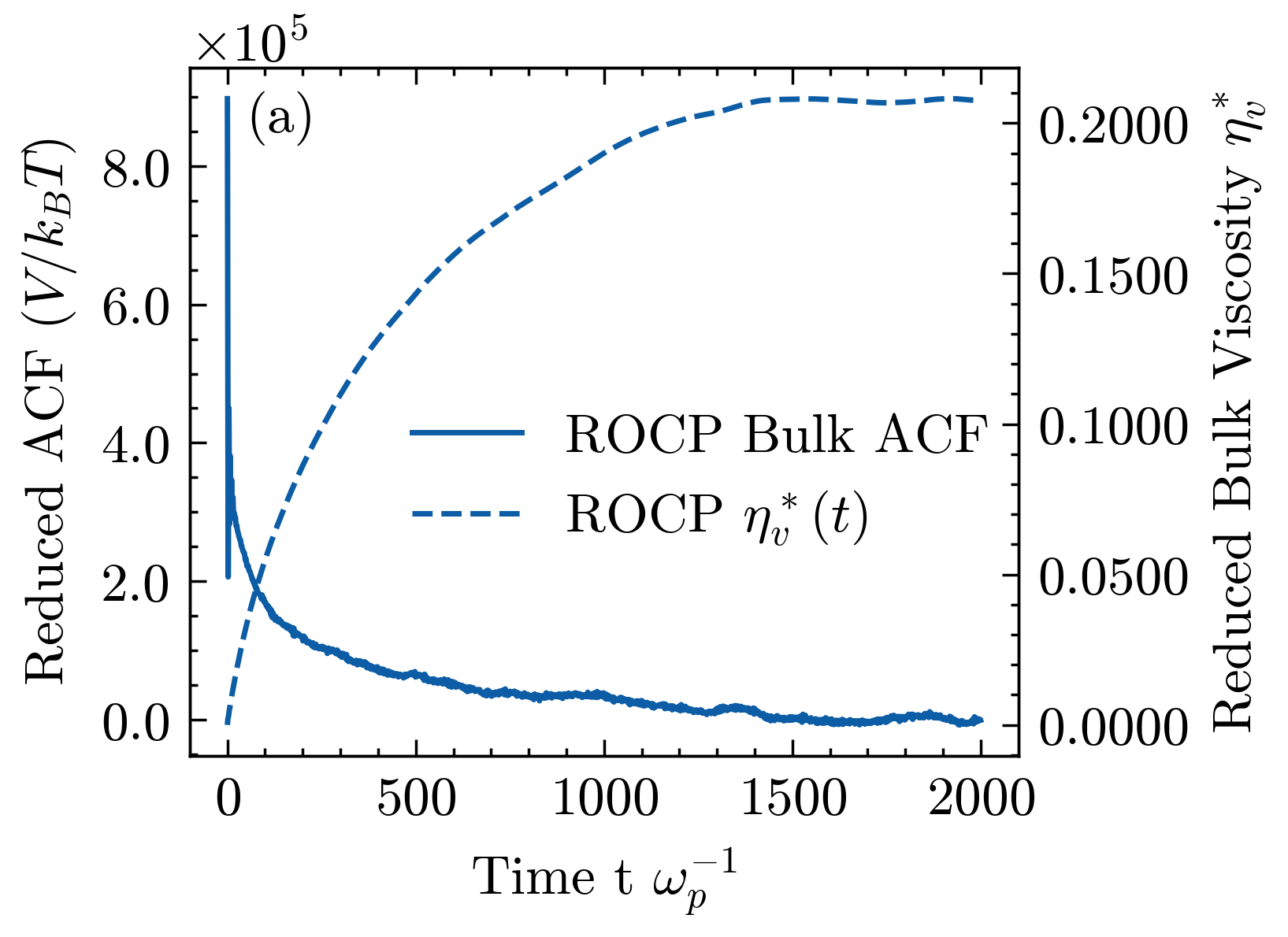}
    
    \includegraphics[width=0.48\textwidth]{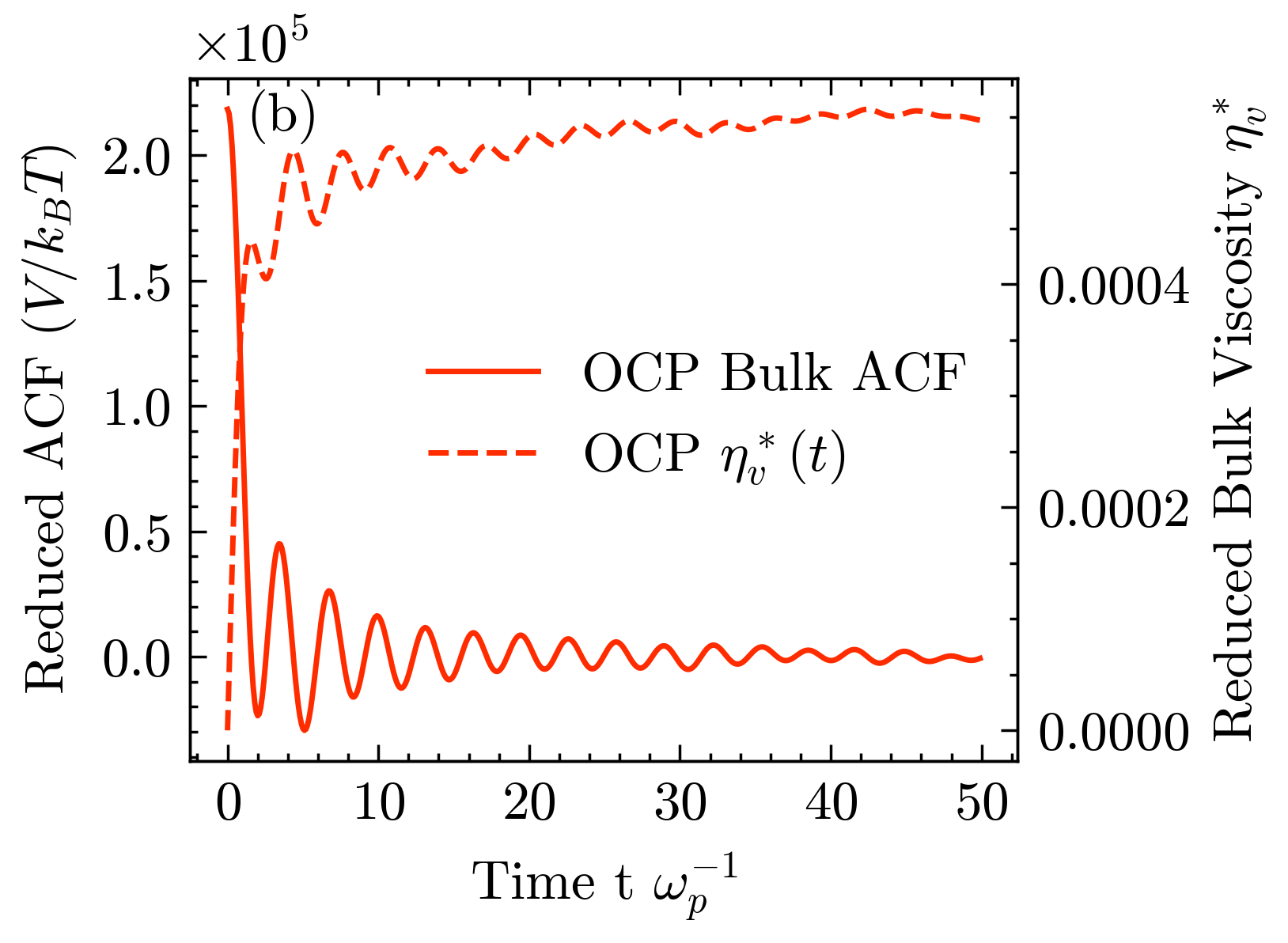}
    \includegraphics[width=0.48\textwidth]{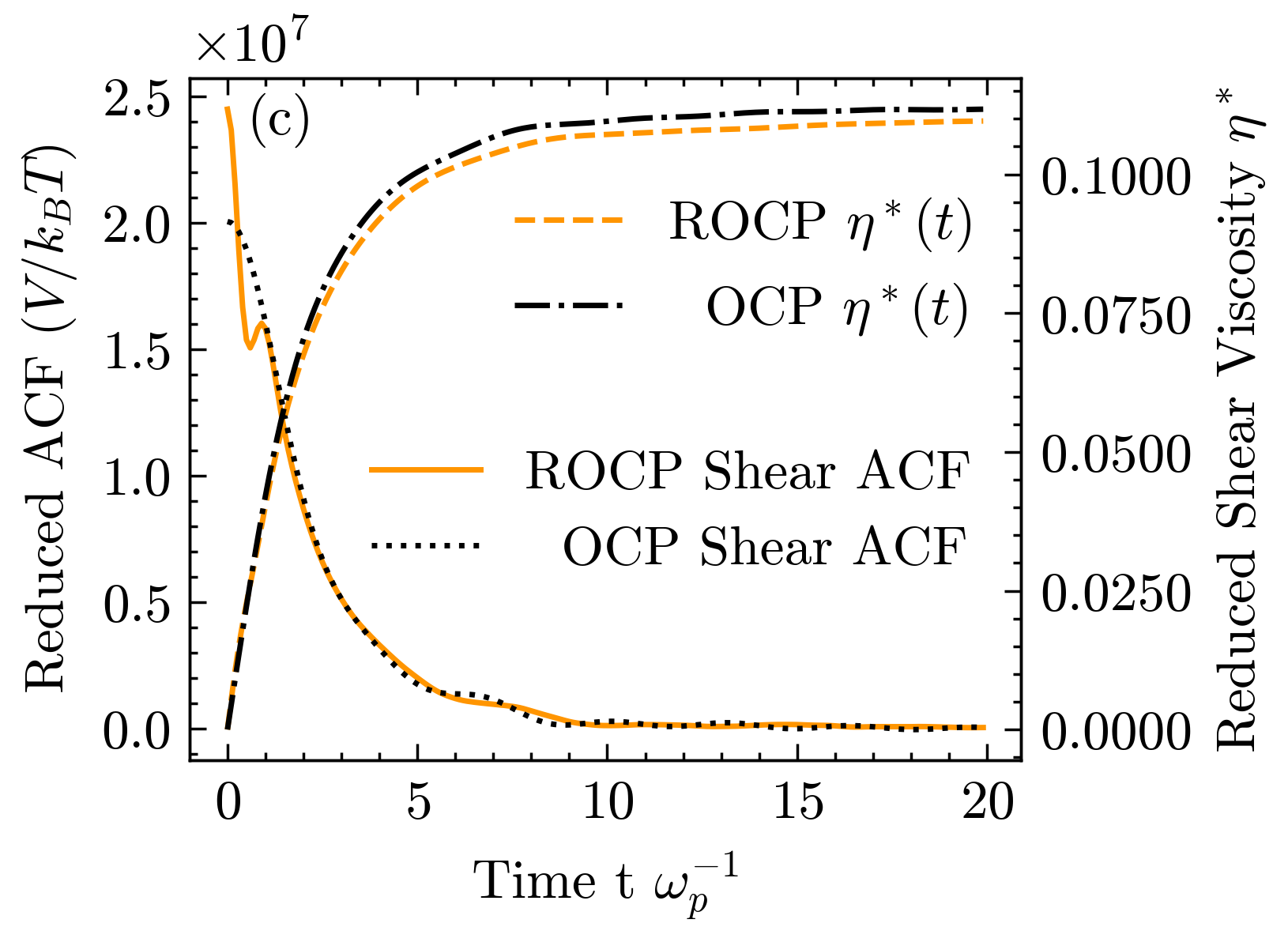}

    \caption{Autocorrelation functions (ACFs) and their respective integrals for (a) ROCP bulk viscosity, (b) OCP bulk viscosity, and (c) ROCP and OCP shear viscosity at $\Gamma = 50$ and $\Omega = 0.14$.} 
    \label{fig:bulk-sim}
\end{figure}

The results of an ROCP bulk viscosity calculation from a particular simulation ($\Gamma = 50$ and $\Omega = 0.14$) is shown in Fig.~\ref{fig:bulk-sim}(a). It can be seen that the pressure autocorrelation function decays over thousands of plasma periods $\sim 2000 \omega_p^{-1}$. This simulation, however, needed to run for $10^{5} \omega_p^{-1}$ to achieve convergence, and still some noise is observed in the decay. As autocorrelation functions get longer, more and more averaging must be done. Several data points required simulation lengths greater than $10^{6} \omega_p^{-1}$.
The long time needed for convergence limited the range for which bulk viscosity values could be computed. Accurate data was obtained for $1 < \Gamma < 100$ and $0.10 < \Omega < 0.30$. 
In physical units, if one takes $r_\textrm{B} \sim 1$~\AA, corresponding to N$_2$, this parameter space corresponds to approximately  $10^{26}$~m$^{-3}$$< n < 10^{28}$~m$^{-3}$, and $250$~K $ < T < 20,000$~K.

Simulations were also run of the traditional OCP. The OCP is subject to less noise than the ROCP, so only 5,000 particles were used. These simulations otherwise proceeded in the same fashion as for the ROCP. 

\begin{figure}
    \centering
    \includegraphics[width=.48\textwidth]{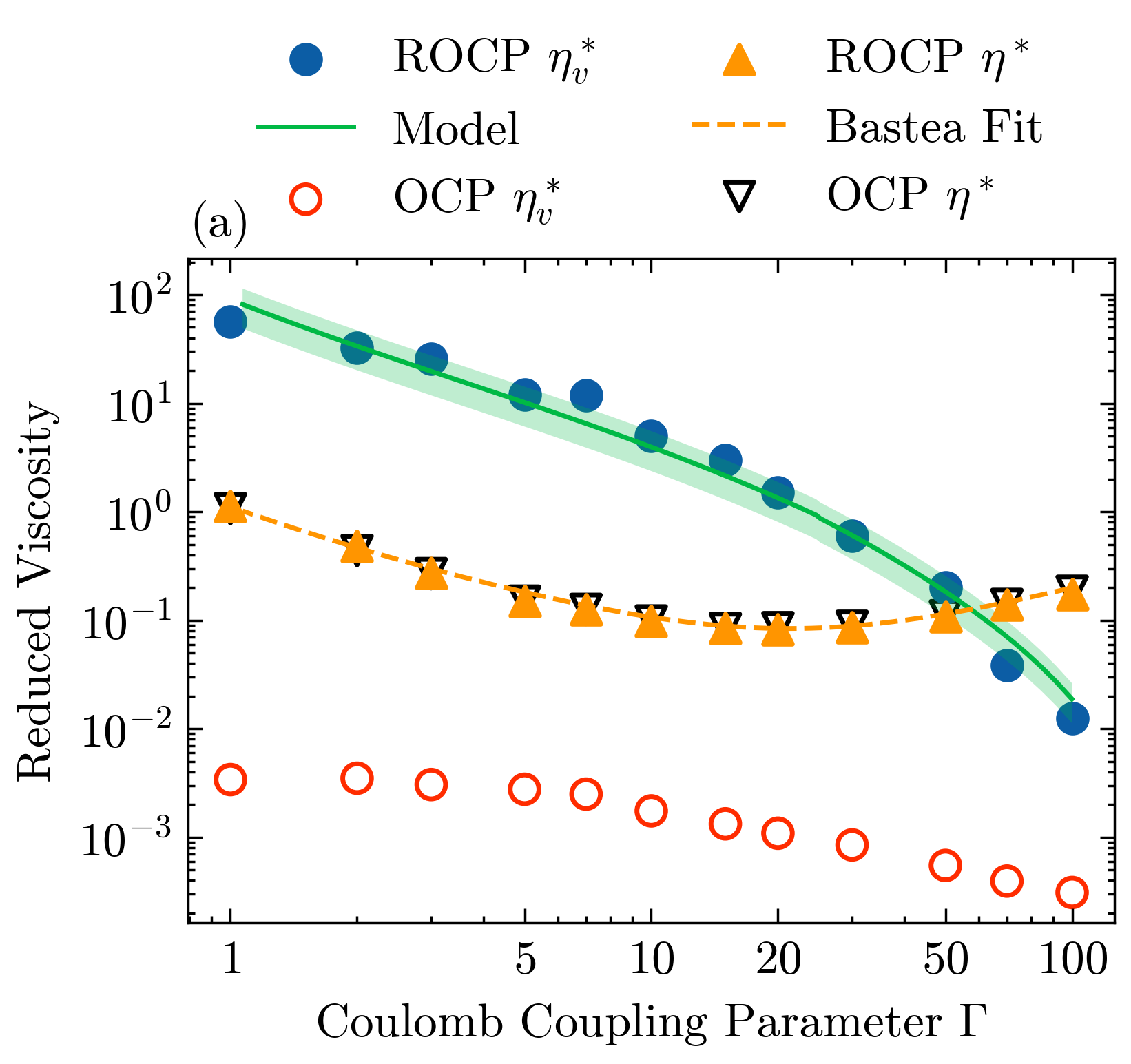}
    \includegraphics[width=.48\textwidth]{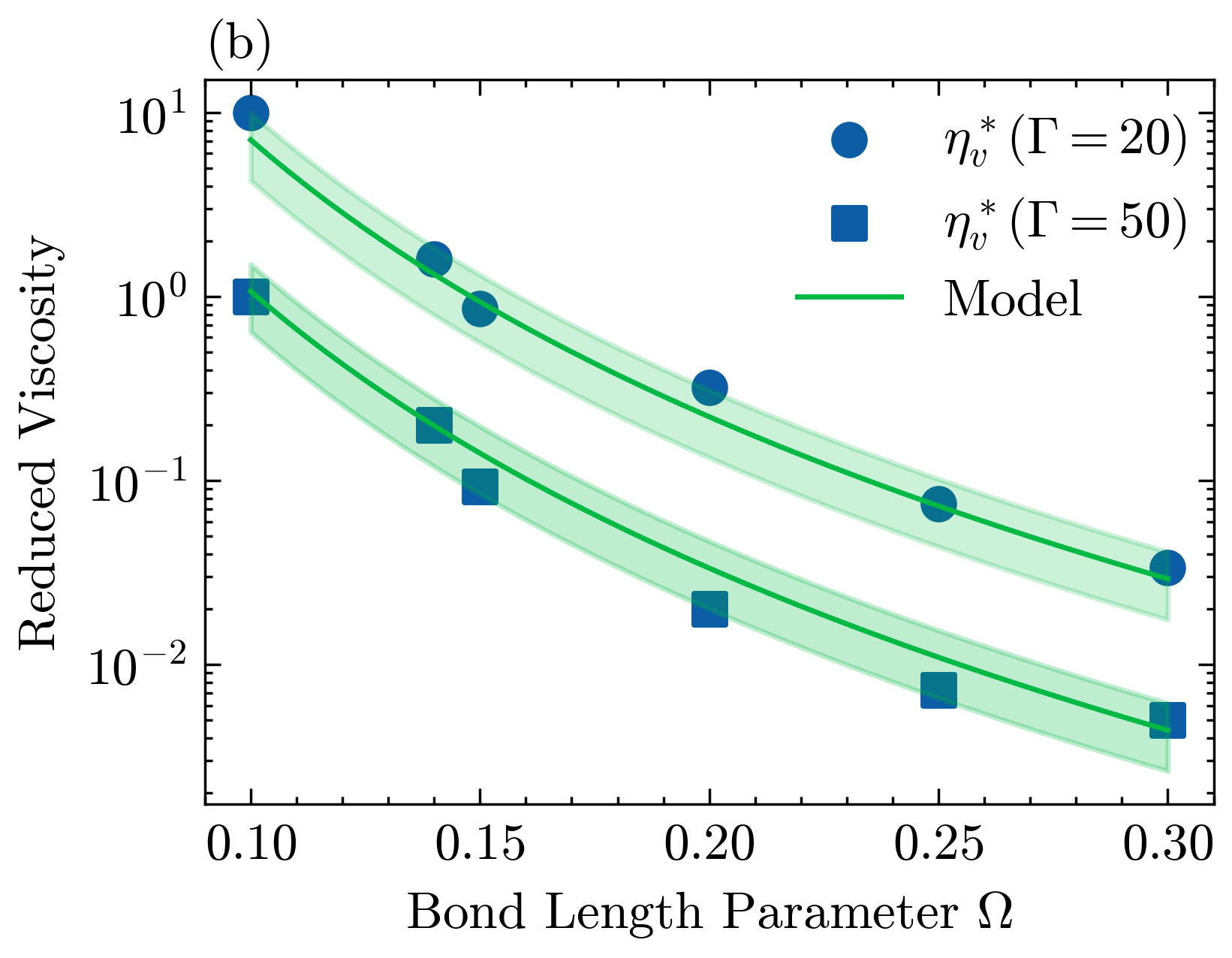}
    \caption{Reduced bulk viscosity variation with (a) the coupling parameter $\Gamma$ and (b) the bond length parameter $\Omega$. In (a), $\Omega$ was held constant at $\Omega = 0.14$. The green region (``Model'') refers to Eq.~(\ref{eq-final-bulk}).}
    \label{fig:bulk-prediction}
\end{figure}

\subsection{Results}
Simulation results are shown in Fig.~\ref{fig:bulk-prediction}, and the data is provided in Tables~\ref{tab:tab1} and~\ref{tab:tab2}. Figure~\ref{fig:bulk-prediction}(a) shows shear and bulk viscosity for both the OCP and ROCP as a function of $\Gamma$ with a constant value of $\Omega = 0.14$ chosen for the ROCP. Figure~\ref{fig:bulk-prediction}(b) shows bulk viscosity of the ROCP as a function of $\Omega$ for fixed values of $\Gamma$ of 20 and 50. 
Values are presented in terms of reduced bulk and shear viscosity, defined as 
\begin{equation}
    \eta_v^* = \frac{\eta_v}{mna^2\omega_p} \quad \textrm{and} \quad \eta^* = \frac{\eta}{mna^2\omega_p} , \label{eq-reduced-bulk-def}
\end{equation}
which correspond to the natural units defined in Sec.~\ref{sec:rocp}. For the ROCP, $m$ is the molecular mass, whereas for the OCP it is the atomic mass.

\begin{table}[]
    \centering
    \begin{ruledtabular}
    \begin{tabular}{ccccc}
        & \multicolumn{2}{c}{OCP} & \multicolumn{2}{c}{ROCP}\\
        $\Gamma$& \multicolumn{1}{c}{$\eta^*$} & $\eta_v^*$ & \multicolumn{1}{c}{$\eta^*$} & \multicolumn{1}{c}{$\eta_v^*$} \\ \hline
        1 & 1.07 & $3.40 \times 10^{-3}$ & 1.12 & 57.0 \\
        2 & 0.440 & $3.50 \times 10^{-3}$ & 0.480 & 32.5 \\
        3 & 0.269 & $3.06 \times 10^{-3}$ & 0.272 & 25.8 \\
        5 & 0.150 & $2.77 \times 10^{-3}$ & 0.149 & 12.0 \\
        7 & 0.126 & $2.50 \times 10^{-3}$ & 0.125 & 11.8 \\
        10 & 0.0985 & $1.75 \times 10^{-3}$ & 0.0980 & 5.00 \\
        15 & 0.0840 & $1.33 \times 10^{-3}$ & 0.0850 & 3.00 \\
        20 & 0.0855 & $1.09 \times 10^{-3}$ & 0.0820 & 1.60 \\
        30 & 0.0880 & $8.50 \times 10^{-4}$ & 0.0860 & 0.600\\
        50 & 0.110 & $5.50 \times 10^{-4}$ & 0.109 & 0.200\\
        70 & 0.140 & $3.95 \times 10^{-4}$ & 0.140 & 0.0385\\
        100 & 0.185 & $3.1 \times 10^{-4}$ & 0.173 & 0.0125\\ 
    \end{tabular}
    \end{ruledtabular}
    \caption{MD results of reduced shear and bulk viscosity of the OCP and ROCP as a function of $\Gamma$. ROCP values use $\Omega = 0.14$.}
    \label{tab:tab1}
\end{table}
\begin{table}[]
    \centering
    \begin{ruledtabular}
       \begin{tabular}{cccc}
      $\Gamma$& $\Omega$ & $\eta^*$ & $\eta_v^*$ \\ \hline
    20 & 0.10 & 0.100 & 10.0 \\
     & 0.14 & & 1.60  \\
     & 0.15 & & 0.864  \\
      & 0.20 & & 0.320  \\
      & 0.25 & & 0.0750  \\
      & 0.30 & & 0.0336  \\ \hline
     50& 0.10 &0.109& 1.00  \\
      & 0.14 & &0.200  \\
      & 0.15 & &0.0925 \\
       & 0.20 & &0.0200 \\
      & 0.25 & & 0.00730  \\
      & 0.30 & & 0.00500 \\
\end{tabular} 
    \end{ruledtabular}
\caption{MD results of reduced shear and bulk viscosity of the ROCP as a function of $\Omega$. Shear viscosity is constant with $
\Omega$.}
    \label{tab:tab2}
\end{table}
The bulk viscosity of the ROCP is found to be larger than the bulk viscosity of the OCP by at least an order of magnitude. This is because the ROCP has two internal degrees of freedom (molecular rotation) which can temporarily trap energy. When the equipartition theorem is violated during an expansion or compression process, equilibrium is not restored until rotational relaxation occurs, leading to a large bulk viscosity. The OCP has no internal degrees of freedom, and hence the time-lag to equilibrium occurs due to structural rearrangement, which is a much faster process than rotational relaxation. This leads to a very small bulk viscosity for the OCP. 

Further insight can be obtained by analyzing the autocorrelation functions shown in Fig.~\ref{fig:bulk-sim}. The pressure autocorrelation function of the OCP decays to zero on the timescale of a single plasma period, implying that this system will restore equilibrium in $\sim \omega_p^{-1}$ following a small perturbation. The autocorrelation function of the ROCP shows a similar decay at early times, demonstrating the mechanism of structural rearrangement is present in the ROCP as well. However, this autocorrelation function does not reach zero until approximately $2000~\omega_p^{-1}$ have passed, as energy is trapped in the two rotational modes.  

Figure~\ref{fig:bulk-prediction} also highlights how bulk viscosity of the ROCP varies with $\Omega$ and $\Gamma$. It can be observed that $\eta_v^*$ increases with decreasing $\Gamma$ and decreasing $\Omega$, with the latter dependence appearing more severe. This scaling can be understood in the context of the rotational relaxation dynamics, so discussion of this physics will be reserved for Sec.~\ref{sec-rot}. 

Figure~\ref{fig:bulk-prediction} shows that the shear viscosity of the ROCP is identical to that of the OCP. This is not a surprising result, for shear viscosity originates from translational motion and center of mass motion in the ROCP is essentially the same as translational motion in the OCP. However, in Fig.~\ref{fig:bulk-sim}, one can see that the early time behavior of their autocorrelation functions differ. The ROCP starts at a larger value and experiences a bump on the path to zero, while the OCP shows a smooth decay. The bump occurs after a time of approximately $\omega_T^{-1} = \sqrt{I/2k_BT}$ has passed, which is the average time it takes for a rigid rotor to rotate. In this sense, it does seem that molecular rotation has an effect on early-time shear stresses. However, the integrated value is essentially the same, so the influence of this on the shear viscosity coefficient is negligible.  
Importantly, it is also shown in Fig.~\ref{fig:bulk-prediction} that the bulk viscosity of the ROCP can exceed the shear viscosity by more than an order of magnitude. In general, when $\eta_v / \eta > 1$, bulk viscosity effects are expected to be significant. 
For instance, this means that the bulk viscosity determines the kinematic longitudinal viscosity, $b=(\frac{4}{3} \eta + \eta_v)/\rho$, rather than the shear viscosity. 

The data in this section suggests that the rotational degrees of freedoms are responsible for the large bulk viscosity coefficient of the ROCP. 
In the next section, non-equilibrium MD simulations of rotational relaxation are used to give insight into the observed $\Gamma$ and $\Omega$ scaling, and to validate the use of Kustova's formula in plasmas. The results are used to construct a model for bulk viscosity of the ROCP and demonstrate that it agrees well with the results of the Green-Kubo simulations. Since rotational relaxation simulations are computationally cheaper than Green-Kubo, data is acquired for a wider range of $\Omega$ values ($0.01-0.30$) and used to predict the bulk viscosity of diatomic ions across a broader range of conditions.   

\section{Model Based on Rotational Relaxation\label{sec-rot}} 
\subsection{MD Setup}
\begin{figure}
    \centering
     \includegraphics{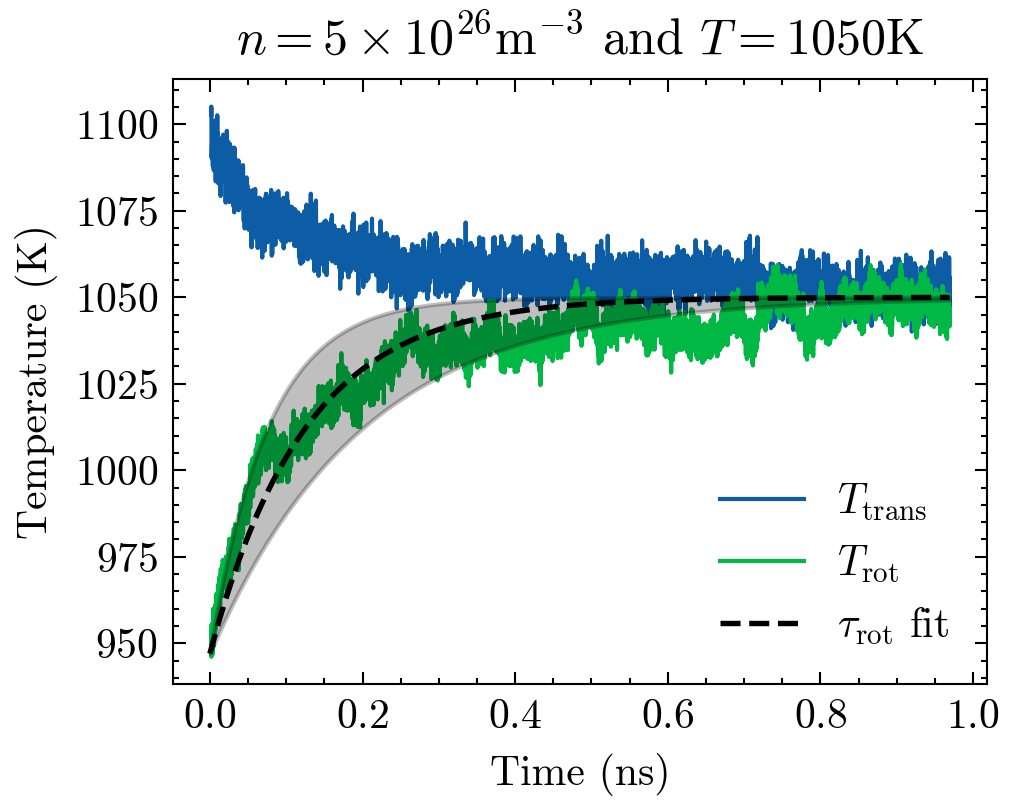}
    \caption{Time-evolution of the translational temperature (blue) and rotational temperature (green) from an MD simulation at the specified density and temperature and $r_\textrm{B} = 1.09$~\AA. The shaded region represents the range of possible curves used to fit the data, with the top half showing better agreement at early times and the bottom half at late times. The dashed line represents the actual curve used to extract $\tau_{\textrm{rot}}$.}
    \label{fig:time-curve}
\end{figure}

Molecular dynamics simulations of rotational relaxation were setup in similar fashion to those from Sec.~\ref{sec-bulk}, with the exception that to start the simulations, the ion translational and rotational degrees of freedom were thermostat to differing temperatures, creating a $\sim 10\%$ temperature perturbation around some desired equilibrium temperature. During the NVE stage that follows, the translational and rotational temperatures come to an equilibrium. An example temperature relaxation curve generated from this procedure can be seen in Fig.~\ref{fig:time-curve}. To extract the rotational relaxation time, an exponential is fit to the temperature data, as has been done in previous work \cite{Haas_1994, Valentini_2012}. 
Since there is some noise in the data, a range of exponentials could be taken as the line of best fit depending on at which times best agreement is prioritized. The shaded region represents this range, which for all data collected was roughly $\pm 40\%$ from the average. This is taken to be the statistical error in the subsequent plots. 

Here, a parameter space relevant to atmospheric pressure plasmas is explored~\cite{LeVan_2024}. In particular, densities from $n = 10^{25}-10^{27}~\textrm{m}^{-3}$, temperatures from $T = 250-5000~\textrm{K}$, and bond lengths between $r_\textrm{B} = 0.545-1.635~\textrm{\AA}$ are simulated. In dimensionless units, this corresponds to $\Gamma = 1-100$ and $\Omega = 0.01-0.30$. 

\subsection{Results}

\begin{figure*}
    \centering
    \includegraphics[width=0.32\textwidth]{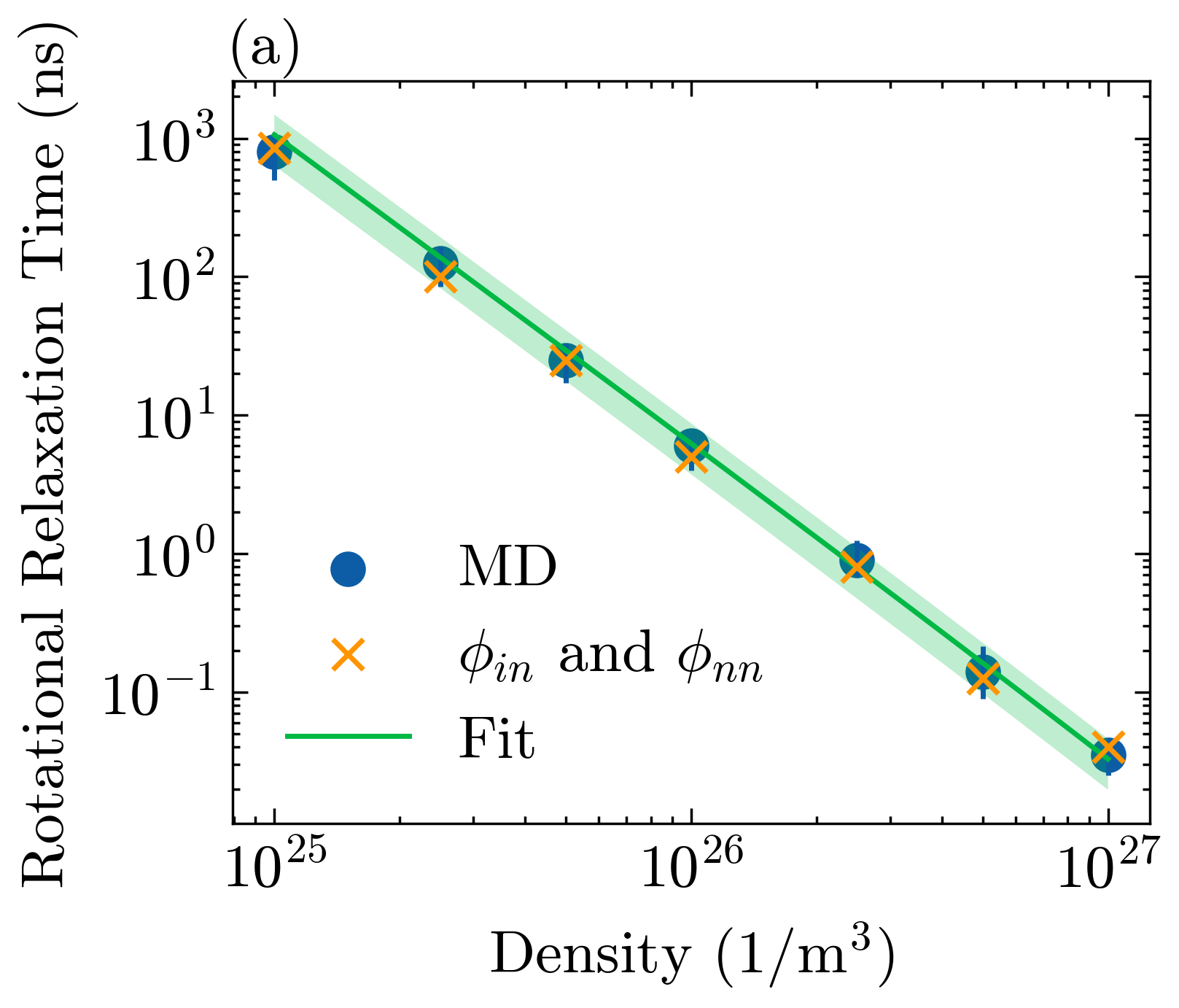}
    \includegraphics[width=0.32\textwidth]{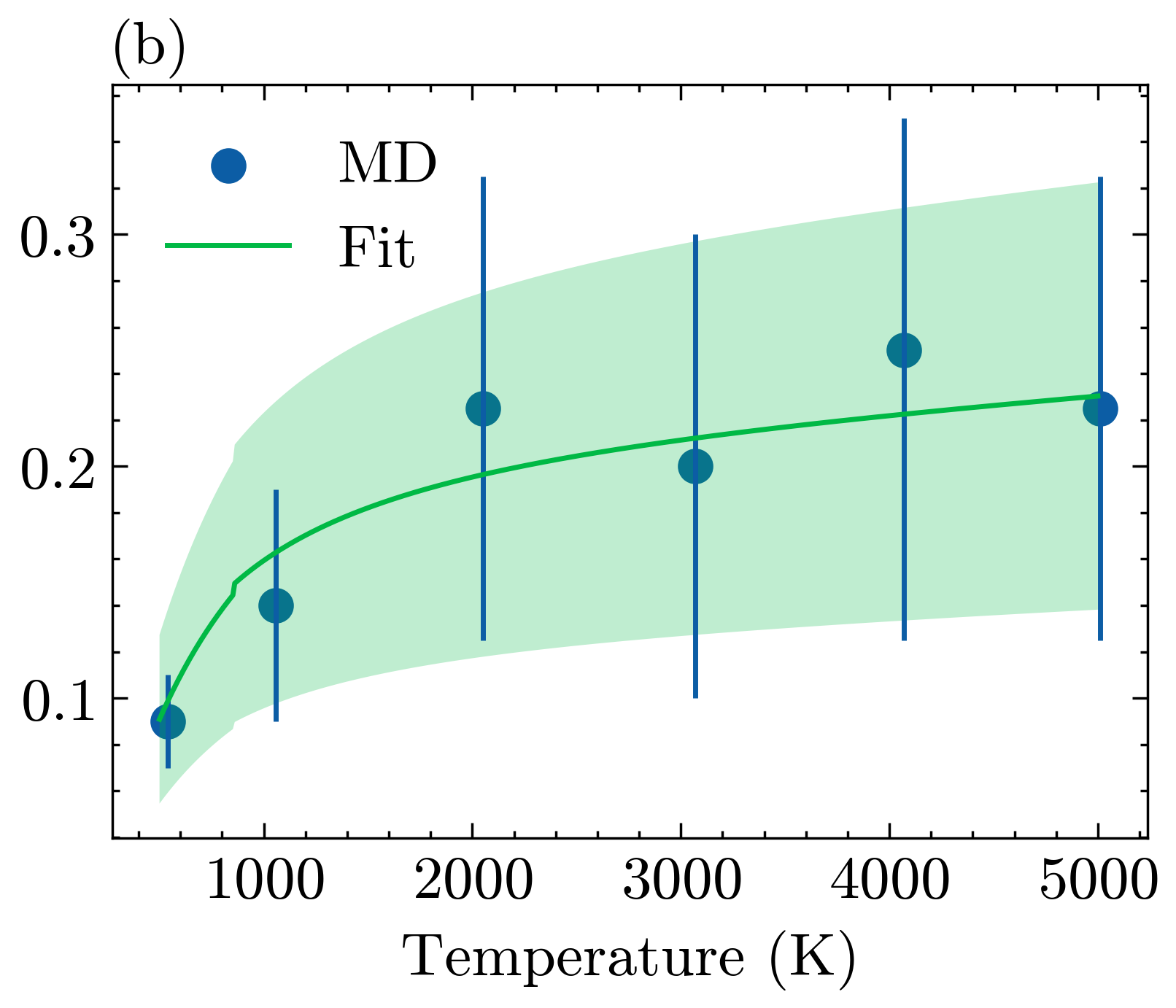}
    \includegraphics[width=0.32\textwidth]{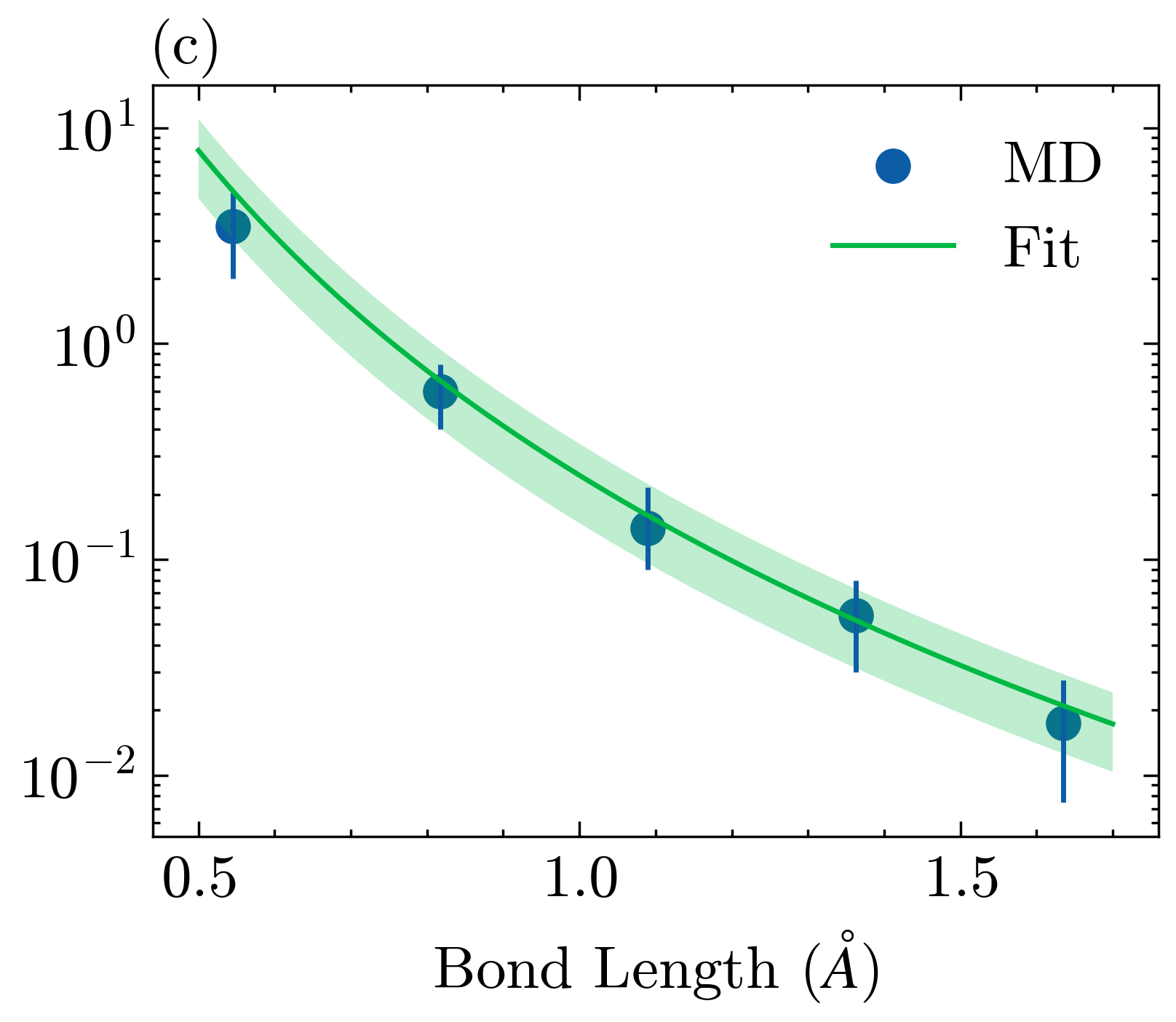}
    \caption{Variation of the rotational relaxation time, $\tau_{\textrm{rot}}$, with (a) density, (b) temperature, and (c) bond length. ROCP data from MD is plotted with blue points and data with ion-neutral and neutral-neutral interactions turned on with orange crosses. The ``fit'' line refers to Eq.~(\ref{eq:rot-fit}) in real units ($\tau_{\textrm{rot}} = \tau_{\textrm{rot}}^*/\nu$) and the shaded region around it represents the uncertainty in the fit ($\pm 40\%$) due to noise in the MD data.
    Aside from when it is the parameter being varied, data corresponds to $n = 5 \times 10^{26}$~m$^{-3}$, $T = 1000$~K and $r_\mathrm{B} = 1.09$~\AA.}
    \label{fig:all-data}
\end{figure*}

A subset of simulation results can be seen in Fig. \ref{fig:all-data}. Error bars on the MD data are generated by looking at the minimum and maximum for $\tau_{\textrm{rot}}$ that still give reasonable agreement with the temperature data. The data indicates that rotational relaxation speeds up with increasing density and bond length, and slows down with increasing temperature. 

We model the data in terms of the reduced rotational relaxation time, defined as $\tau^*_{\textrm{rot}} = \tau_{\textrm{rot}} \nu$ where $\nu$ is the collision frequency. It should be noted that $\tau^*_{\textrm{rot}}$ is defined identically to the rotational collision number commonly used in studies of neutral gas, $Z = \tau_{\textrm{rot}} \nu$, which represents the number of collisions required to reach equilibrium. However, in strongly coupled plasmas, collisions are not binary, so the collision frequency $\nu$ only represents the frequency at which a particle deviates 90 degrees from its original trajectory. For this reason, we will not adopt the $Z$ convention.

In the OCP, $\nu$ has been studied in the context of diffusion~\cite{Daligault_2012, glasstone1941theory}. Though diffusion is a different physical process from viscosity and rotational relaxation, the collision frequency found in these works is still applicable because it scales with $\Gamma$ in a way that captures strong coupling effects. 
From these works on diffusion, $\nu$ can be written as~\cite{Daligault_2012, glasstone1941theory}
\begin{eqnarray}
    \nu = 
    \left\lbrace \begin{array}{ll} 
    \alpha \nu_0 , & \textrm{for $\Gamma < 25$} \label{eq-coll-freq} \\
    \dfrac{\Gamma}{A}e^{B \Gamma} \dfrac{k_\textrm{B} T}{m a^2 \omega_p} & \textrm{for $\Gamma > 25$} ,
    \label{eq-coll-freq-2}
    \end{array} \right.
\end{eqnarray}
For the $\Gamma < 25$ equation, $\alpha = 0.647$, 
\begin{equation}
    \nu_0 = \frac{4n}{3} \sqrt{\frac{\pi}{m}} \biggl(\frac{e^2}{4 \pi \epsilon_0} \biggr)^{2} \frac{\ln \Lambda}{(k_\textrm{B} T)^{3/2}}
\end{equation}
is the collision frequency predicted in weak coupling predicted by Landau-Spitzer  theory, and
\begin{equation}
\label{eq-ln_lambda}
    \ln \Lambda = \ln \left( 1 + C \frac{\lambda_D}{r_L}\right)
\end{equation}
is a modified Coulomb logarithm that extends the Landau-Spitzer expression for $\nu$ to $\Gamma \approx 25$~\cite{Daligault_2012}. Here, $C = 2.159$ is a fit factor, $\lambda_\textrm{D} = \sqrt{\epsilon_0 k_\textrm{B} T /e^2 n}$ is the Debye length, and $r_\textrm{L} = e^2 /4 \pi \epsilon_0 k_\textrm{B} T$ is the Landau length (distance of closest approach). For the $\Gamma > 25$ equation, $A = 1.52$ and $B = 0.0082$.

By fitting to the relaxation data, we obtained the following expression 
\begin{equation}
    \tau^*_{\textrm{rot}} = \tau_{\textrm{rot}} \nu = \alpha_0 e^{-\alpha_0 \Gamma} \Omega^{-5}
    \label{eq:rot-fit}
\end{equation}
where $\alpha_0 = 0.18$. 
There are several important physical effects to extract from this relationship. For one, rotational relaxation grows longer with decreasing $\Gamma$ and $\Omega$, consistent with the findings in the previous section. This matches physical intuition as well; as $\Gamma$ decreases, i.e., the ratio of interaction energy to kinetic energy decreases, forces on the molecules become weaker and hence energy exchange between translational and rotational modes takes longer. And as $\Omega$ decreases, i.e., the ratio of the bond length to average inter-particle spacing decreases, molecules begin to look more like atoms, effectively shielding the rotational degrees of freedom. 

However, the scaling is in sharp contrast to what is predicted in neutral gas, implying the presence of different physics processes in plasmas. For instance, Parker's model, a standard formula for estimating rotational relaxation in diatomic neutral gas, predicts $\sim r_\textrm{B}^{-2}$ scaling with bond length and no scaling with density outside of the translational collision frequency~\cite{Parker_1959}. Eq.~(\ref{eq:rot-fit}), on the other hand, demonstrates a $r_\textrm{B}^{-5}$ scaling with bond length and exponential scaling of density beyond the collision frequency. 

We propose two mechanisms unique to plasmas that are likely responsible for this difference. For one, at STP density and temperature, neutral gas is considered dilute, as collisions between molecules are relatively sparse and binary. However, a strongly coupled plasma at the same density and temperature is considered dense, as many-body effects are present due to the long-range nature of the Coulomb force. Many-body effects likely alter the rotational relaxation dynamics, thereby causing a deviation from neutral gas behavior. 

For another, in the ROCP only one atom per molecule interacts, making activation of rotational degrees of freedom in a collision more difficult. This effect, however, is not merely an artifact of ignoring ion-neutral and neutral-neutral interactions in the ROCP, but rather a result of the Coulomb potential being far stronger than the ion-neutral or neutral-neutral interaction potentials at the location of the interaction. Due to their long-range nature, ion-ion interactions effectively shield the other interactions, which is why they are left out of the ROCP. To confirm that this is a good approximation, MD simulations were run with ion-neutral and neutral-neutral interactions turned on, using values corresponding to Nitrogen. Neutral-neutral interactions were modelled with a Lennard-Jones potential and ion-neutral interactions with a charge-induced dipole potential with an artificial repulsive core to prevent bound states, the details of which can be found in previous work~\cite{LeVan_2024}. The results of these simulations are shown in Fig.~\ref{fig:all-data}(a). It can be seen that adding ion-neutral and neutral-neutral interactions has no effect on the rotational relaxation time, confirming intuition.

In comparison to a neutral gas subject to the same conditions, the ROCP has a very long rotational relaxation time. At a temperature of 1000~K and density $2.5 \times 10^{25}$~m$^{-3}$ (atmospheric pressure), MD simulations revealed $\tau_{\textrm{rot}} = 125~\textrm{ns}$ for the ROCP. At the same temperature and density, MD simulations of neutral N$_2$ gas ran by Valentini \textit{et al}~\cite{Valentini_2012} found $\tau_{\textrm{rot}} \approx 0.5~\textrm{ns}$. The shielding of neutral interactions discussed in the previous paragraph is likely responsible, as with only one interacting atom, molecules in the ROCP cannot exchange translational and rotational energy as efficiently.

\subsection{Predicting Bulk Viscosity}

Results of the rotational relaxation simulations can be used to predict bulk viscosity using Kustova's formula. Plugging the rotational relaxation fit formula from Eq.~(\ref{eq:rot-fit}) into the ROCP expression for bulk viscosity from Eq.~(\ref{eq-simple}) provides 
\begin{eqnarray}
    \frac{1}{\eta_v^*} = 
    \left\lbrace \begin{array}{ll}
    \sqrt{\dfrac{3}{\pi}} \dfrac{\alpha}{\alpha_0} e^{\alpha_0 \Gamma} \Gamma^{5/2} \Omega^{5} \left(\dfrac{5}{2} + c_{v,x}\right)^{2} \ln \Lambda, & \Gamma < 25
    \label{eq-final-bulk} \\
    \dfrac{e^{(B+\alpha_0)\Gamma}}{A \alpha_0} \Gamma \Omega^{5} \left( \dfrac{5}{2} + c_{v,x}\right)^2,  & \Gamma > 25. 
    \end{array} \right. 
\end{eqnarray}
Here, $\alpha$ is the constant from the translational collision frequency formula from Eq.~(\ref{eq-coll-freq}), $\alpha_0$ is a constant from rotational relaxation Eq.~(\ref{eq:rot-fit}), and $\ln \Lambda$ is the modified Coulomb logarithm from Eq.~(\ref{eq-ln_lambda}). 
In the second term, $A$ and $B$ are the constants for the collision frequency from Eq.~(\ref{eq-coll-freq-2}). 

The result of Eq.~(\ref{eq-final-bulk}) is shown in Fig.~\ref{fig:bulk-prediction}. This shows very good agreement with data computed from the Green-Kubo relation, validating the use of Kustova's formula in the ROCP. We can therefore use Eq.~(\ref{eq-final-bulk}) to estimate the bulk viscosity of the ROCP for $1 < \Gamma < 100$ and $0.01 < \Omega < 0.30$.

\begin{figure}
    \centering
\includegraphics{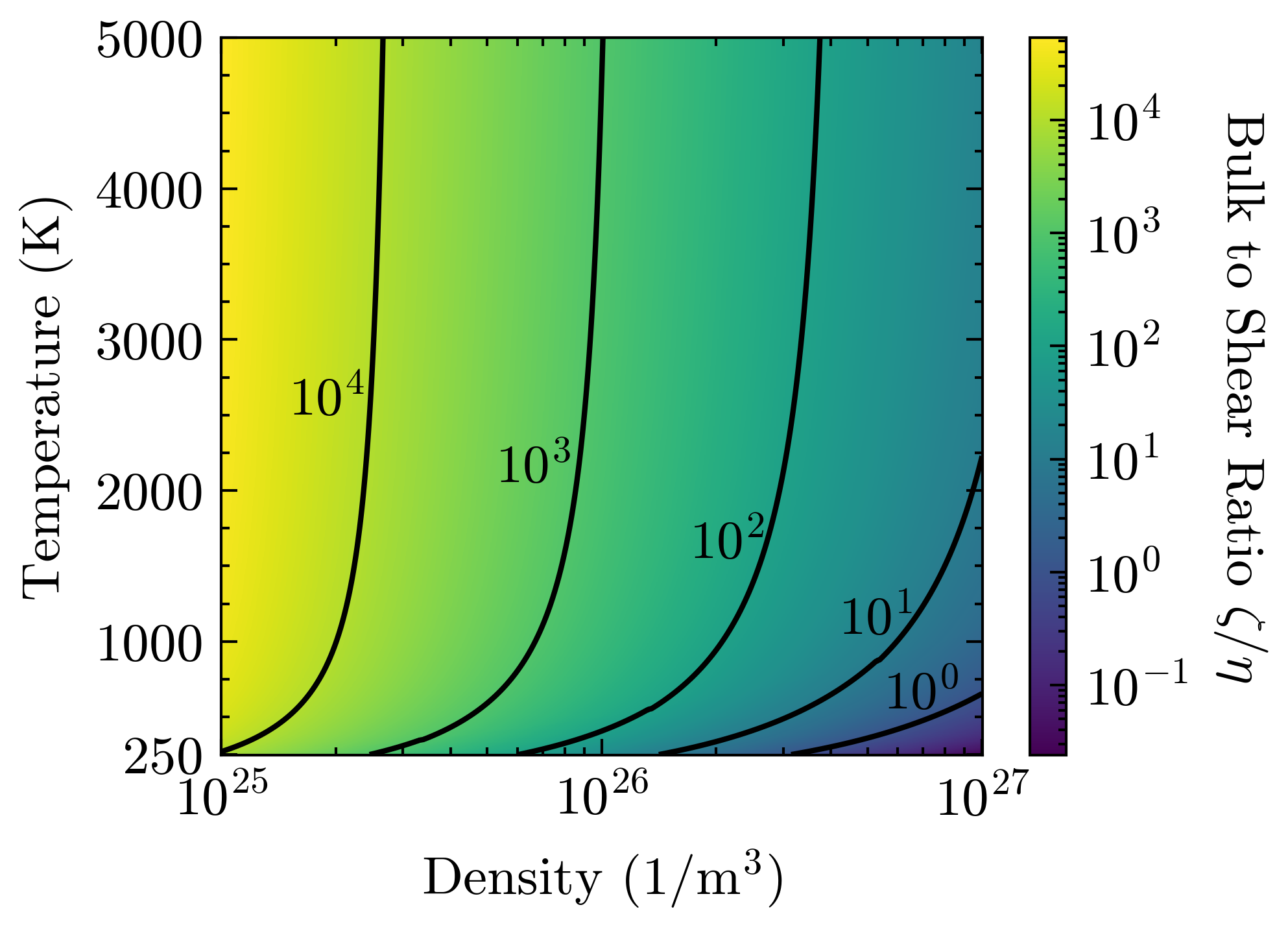}
    \caption{Ratio of bulk to shear viscosity coefficients, computed using the model from Eqs.~(\ref{eq-final-bulk}) and~(\ref{eq-bastea-shear}). Here, a bond length of $r_\textrm{B} = 1.09~\textrm{\AA}$ was used.}
    \label{fig-bulk-to-shear}
\end{figure}

Figure~\ref{fig-bulk-to-shear} shows the bulk to shear ratio across the entire density-temperature space studied (with $r_{\mathrm{B}} = 1.09$~\AA), computed using Eq.~(\ref{eq-final-bulk}) and the following fit to OCP shear viscosity from Bastea~\cite{Bastea_2005}
\begin{equation}
    \eta^* = 0.482\Gamma^{-2} + 0.629\Gamma^{-0.878} + 0.00188\Gamma .
    \label{eq-bastea-shear}
\end{equation}
At a density corresponding to STP ($n = 2.5 \times 10^{25}$~m$^{-3}$), and room temperature, a bulk to shear ratio $\eta_v / \eta \sim 10^3$ is predicted. For context, neutral N$_2$ at atmospheric pressure and 1000~K has a bulk to shear ratio of just $\eta_v / \eta \sim 1$~\cite{Sharma_2022}. This highlights an important point: effects unique to plasmas can lead to large bulk viscosity, motivating a need for further studies in the field of plasma physics. 

\section{Conclusion\label{sec:conclusion}}
In this work, bulk viscosity was computed for systems of strongly coupled diatomic ions using the framework of the rigid rotor one-component plasma. Equilibrium MD simulations were run and the Green-Kubo relations were used to compute the coefficients of shear and bulk viscosity. Results show that the shear viscosity of the OCP and ROCP are identical, but the bulk viscosity of the ROCP is much larger than for the OCP. 
This is attributed to a long relaxation time of the rotational degree of freedom in the ROCP, causing a long time-lag for the mechanical pressure to relax to the thermodynamic pressure. 
This relaxation rate is longer, and therefore the bulk viscosity larger, particularly when $\Gamma$ and $\Omega$ are small.

Previous work has proposed models connecting the rotational relaxation rate and bulk visocisty coefficients in dilute neutral gases. 
Here, these models were applied to the ROCP, but through modified expressions appropriate to the rotational relaxation time for ionized molecules in a strongly coupled regime. 
Non-equilbrium MD simulations were then run to compute the rotational relaxation time. By plugging the results into Kustovas's formula for bulk viscosity, we were able to validate the use of Kustovas's formula in plasmas, and construct a simple model for bulk viscosity of the ROCP. These simulations also gave insight into the underlying physics responsible for the ROCP's large bulk viscosity and the way in which it scales. The long-range nature of the Coulomb force was found to be a significant factor because it shields the ion-neutral and neutral-neutral interaction forces. 

This work concentrated on the ROCP as a reduced model that demonstrates the relevant physics of molecular rotation in ionized gases. However, future work will be required to connect this to physical systems. 
In the ROCP, all particles are singly-ionized diatomic molecules, regardless of the density and temperature conditions. 
Of course, physical systems will commonly consist of some mixture of molecular ion species, and atomic ion species, as well as neutral molecular and atomic species. 
The relative concentrations of each will be determined by a complex combination of reactions, such as ionization, dissociation, recombination, and molecular bonding. 
Exploring the influence of multiple species will be a topic of future work. 

The results presented here demonstrate that phenomena unique to plasmas can affect bulk viscosity in a significant manner, motivating a need for further study of how bulk viscosity influences plasma hydrodynamics. 
For instance, the hydrodynamic sound attenuation coefficient is proportional to the kinematic longitudinal viscosity coefficient $b=(\frac{4}{3} \eta + \eta_v)/\rho$~\cite{Hansen_2013}. 
This suggests that in plasmas with $\eta_v \gg \eta$, bulk viscosity can be the dominant effect determining the damping rate of sound waves. 

There are many other potential applications. 
A recent paper on astrophysical plasmas showed that large bulk viscosity can strongly suppress compressible modes in the turbulent dynamo~\cite{beattie_2023}. 
A number of studies have shown that including bulk viscosity is necessary to agree with experimental data for shock wave structure \cite{Elizarova_2007,Chikitkin_2015,Kosuge_2018}. Emanuel and Argrow showed that in dense molecular gas, shock thickness is roughly linear with the bulk to shear ratio \cite{Emanuel_1994}. Since the bulk to shear ratio of the ROCP can be thousands of times larger than its neutral gas counterpart at the same conditions, assuming neutral gas predictions for bulk viscosity carry over to plasma may lead to signficant under-predictions of shock width. 
Another important implication of a large bulk viscosity is its effect on turbulence. 
It has been shown to increase the decay rate of turbulent kinetic energy and render compressible turbulence incompressible \cite{Pan_Johnsen_2017, Touber_2019, Chen_2019}. 
And for the Rayleigh-Taylor instability, it has been shown to affect the growth of the mixing layer and lead to a more consistent time variation of total entropy \cite{Sengupta_2016}. Each of these are important phenomena in plasmas, further motivating the need for bulk viscosity to be studied in plasma physics.

\begin{acknowledgements}
This research was supported by the US Department
of Energy under Award No. DE-SC0022201. It was also
supported by computational resources and services provided
by Advanced Research Computing (ARC), a division of
Information and Technology Services (ITS) at the University
of Michigan, Ann Arbor.
\end{acknowledgements}

\bibliography{apssamp}

\end{document}